\documentstyle[11pt,newpaspSTV03,twoside,epsf]{article} \def\edcomment#1{\iffalse\marginpar{\raggedright\sl#1\/}\else\relax\fi}
\reversemarginpar

\begin{document}
\title{Synchrotron-Self Compton Spectral Evolution\\ of PKS 2155-304}
 \author{S. Ciprini and G. Tosti}
\affil{- Physics Department \& Astronomical Observatory, Perugia
University\\ - INFN Perugia Section, via Pascoli, 06123, Perugia,
Italy}

\begin{abstract}
The high frequency peaked blazar PKS 2155-304 is one of the
brightest and most intensively studied prototype of BL Lac
objects. Gamma-rays from PKS 2155-304 have been detected from the
MeV to TeV ranges. We computed a synchrotron self-Compton (SSC)
model, based on the temporal behavior of the particles
distribution, responsible for the high-energy emission. Using the
available simultaneous multiwavelength data, we simulated the
overall spectral energy distribution (SED) and the spectral
variability of this source.
\end{abstract}

\section{Introduction}
The overall spectral energy distribution (SED) of blazars show a
two-bump structure, where the lower frequency hump is peaked
either in the IR/optical (low frequency peaked LBL or ``red''
blazar) or in the UV/X-ray bands (high frequency peaked HBL or
``blue'' blazars), and is believed to be produced by synchrotron
emission, while the higher frequency component should be due to
inverse Compton (IC) scattering of soft photons. In the standard
inner-jet scenario, synchrotron radiation and IC scattering were
usually interpreted as due to diffusive shock acceleration of
charged particles within a plasma jet, which itself moves at
relativistic speed and points toward the observer. In this elegant
view, the same relativistic particles up-scatter by IC, the seed
photons produced by synchrotron radiation (synchrotron
Comptonization or synchrotron self-Compton, SSC, process). We
implemented an SSC time-dependent model for the HBL (and
TeV-blazar) PKS 2155-304 (z=0.116). This blazar is a strong
$\gamma$-ray emitter, and one of the more luminous at UV and X-ray
wavelengths. Because of its brightness, it is one of the best
targets of multiwavelenght campaigns, and the data produced in
this campaigns, represent a good test bench to constraint source
models, parameters and simulated SEDs.

\section{Model and simulations}
To constraint our model we used three multiwavelenght campaigns on
PKS 2155-304. The first large campaign, performed in May 1994 (see
Fig. 1), using radio-optical-nearIR ground based observations and
IUE, EUVE, ASCA data, together with the GeV detections obtained in
1994 by EGRET, on board of the Compton Gamma-Ray Observatory
(CGRO) (Urry et al. 1997, Pesce et al. 1997, Pian et al. 1997).
Then we took the data of the May-June 1996 campaign (see Fig. 2),
carried out, in particular, with the infrared (2.8-100\micron)
observation of ISO (Bertone et al. 2000). Moreover we used the
September-November 1997 TeV positively detections by the Cerenkov
Telescope Mark 6 (Narrabri-Australia) and by EGRET (Chadwick et
al. 1999, Vestrand \& Sreekumar 1999, Kataoka et al. 2000,
Chiappetti et al. 1999) (see Fig. 3). This blazar was also
positively detected at MeV regimes by OSSE in 1992 (on board of
CGRO).
\begin{figure}[t]
 \epsfysize=4.5cm \hspace{-0.2cm} \epsfbox{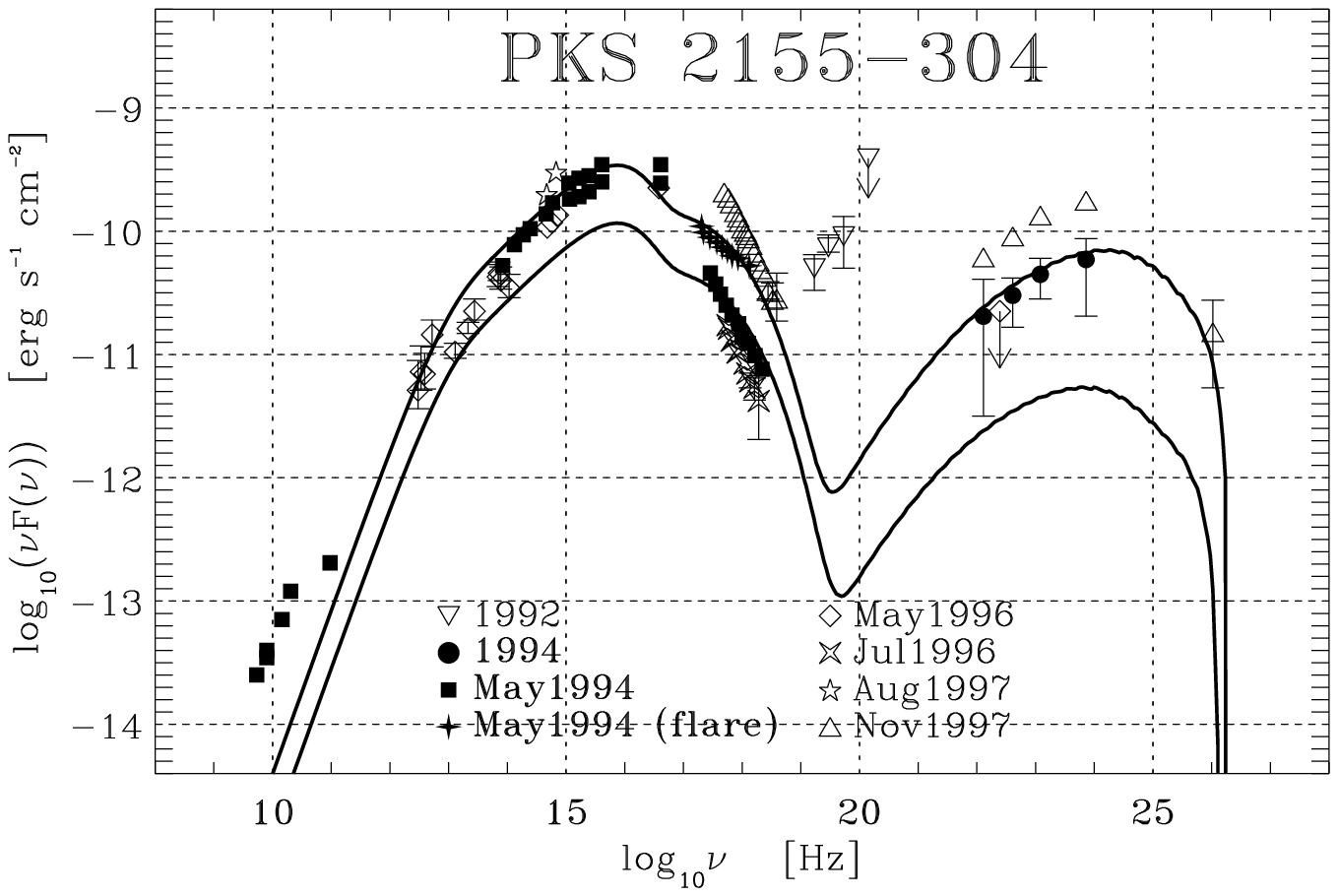}
 \epsfysize=4.5cm \hspace{-0.2cm} \epsfbox{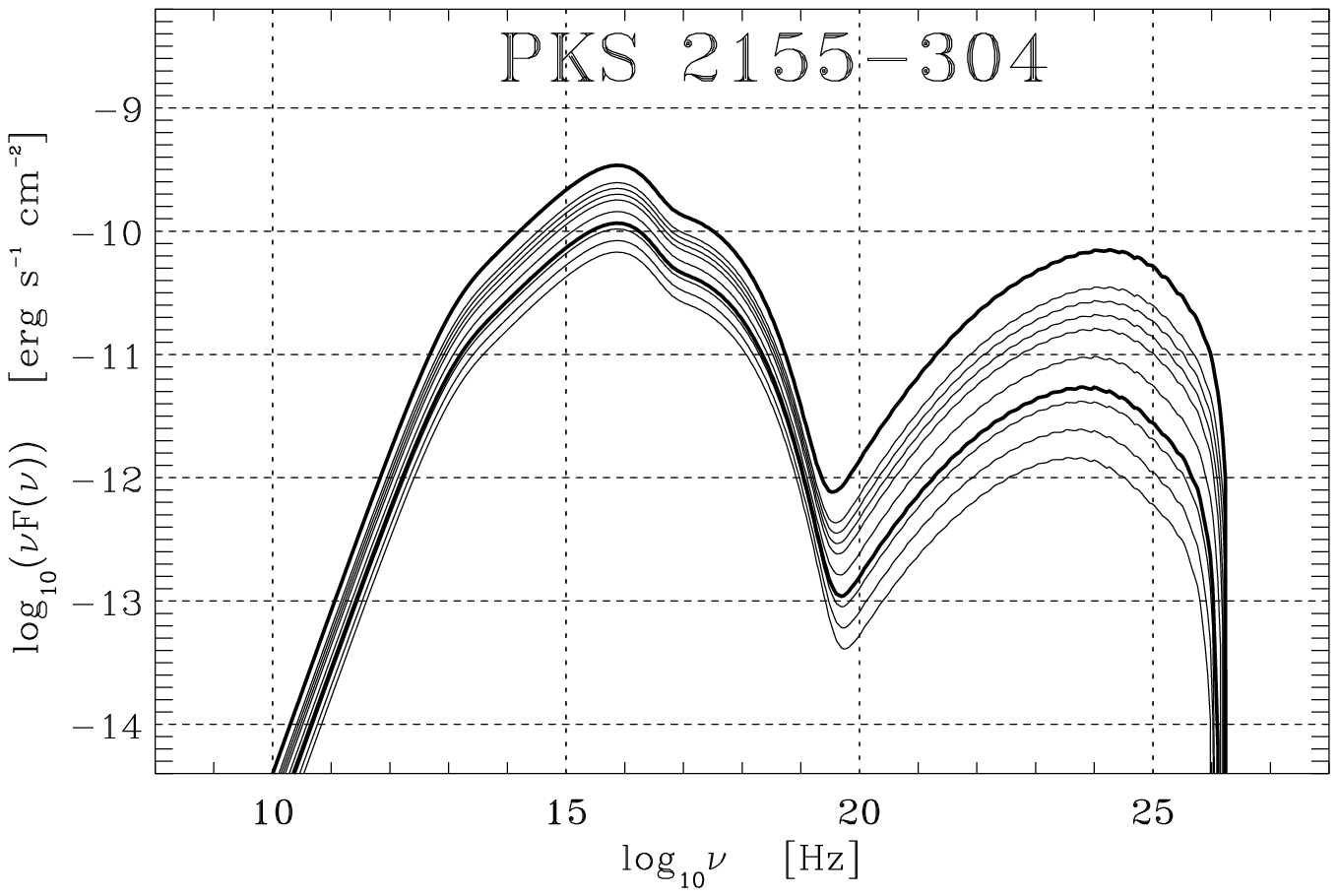}
\caption[h]{Attempts to fit the SED of PKS 2155-304, using the
data of the May 1994 multiwavelength campaign. In the right panel
the temporal evolution.}\label{fig1} \vspace{-2mm}
\end{figure}
%
\begin{figure}[t]
 \epsfysize=4.5cm \hspace{-0.2cm} \epsfbox{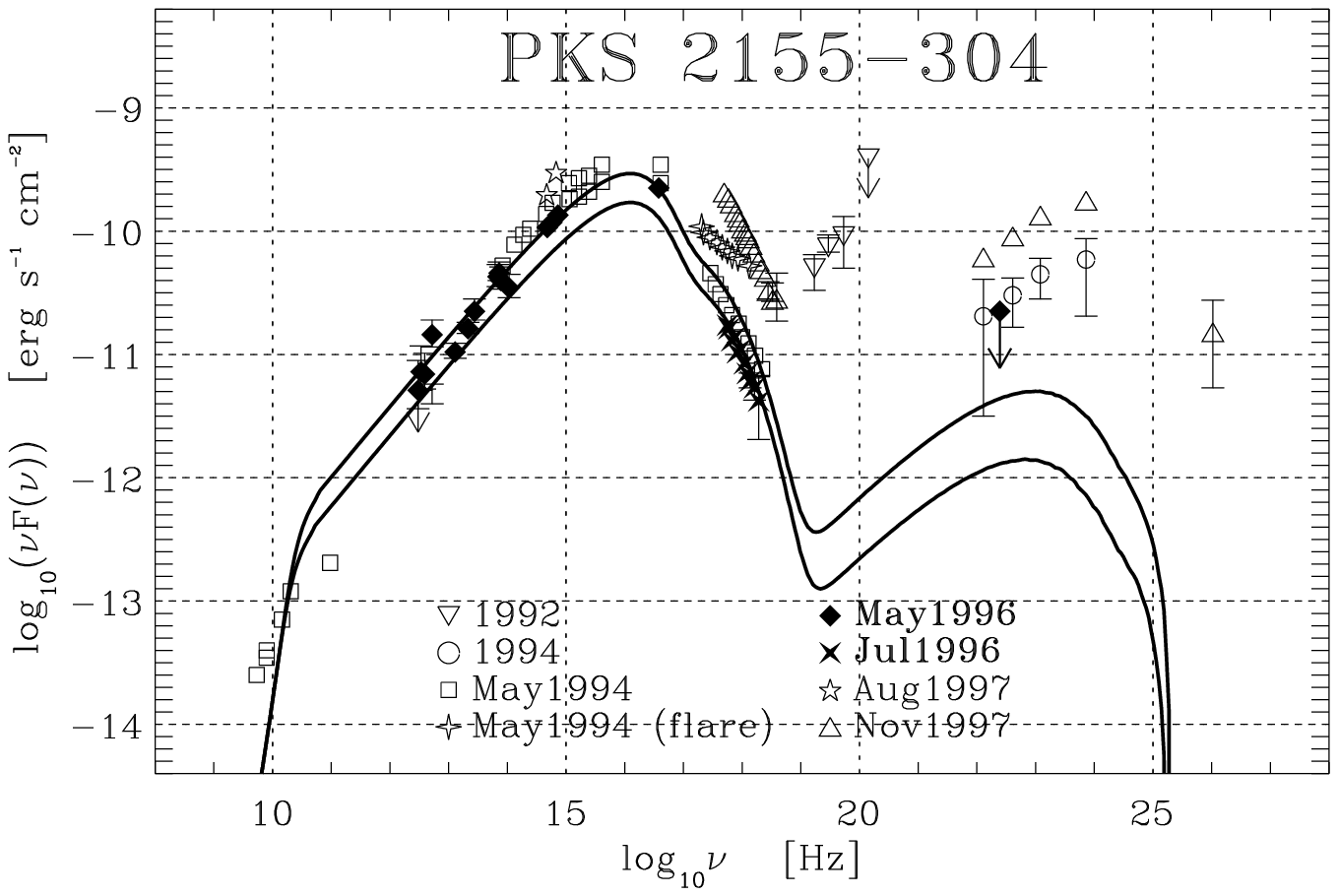}
 \epsfysize=4.5cm \hspace{-0.2cm} \epsfbox{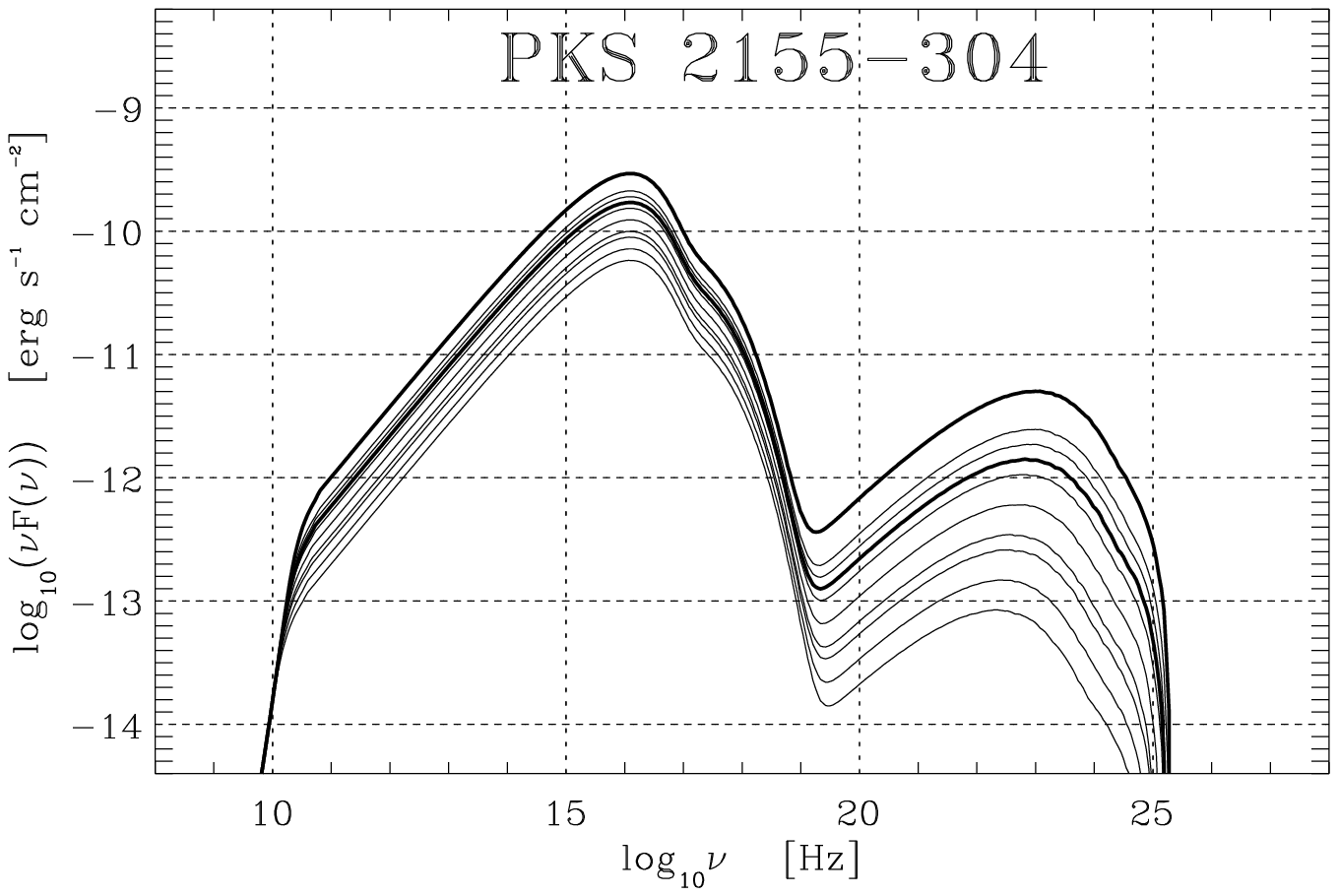}
\caption[h]{Attempts to fit the SED of PKS 2155-304, using the
data of the May 1996 multiwavelength campaign and the July 1996
ASCA data. In the right panel the temporal evolution.}
\label{fig2} \vspace{-2mm}
\end{figure}
%
\begin{figure}[t]
 \epsfysize=4.5cm \hspace{-0.2cm} \epsfbox{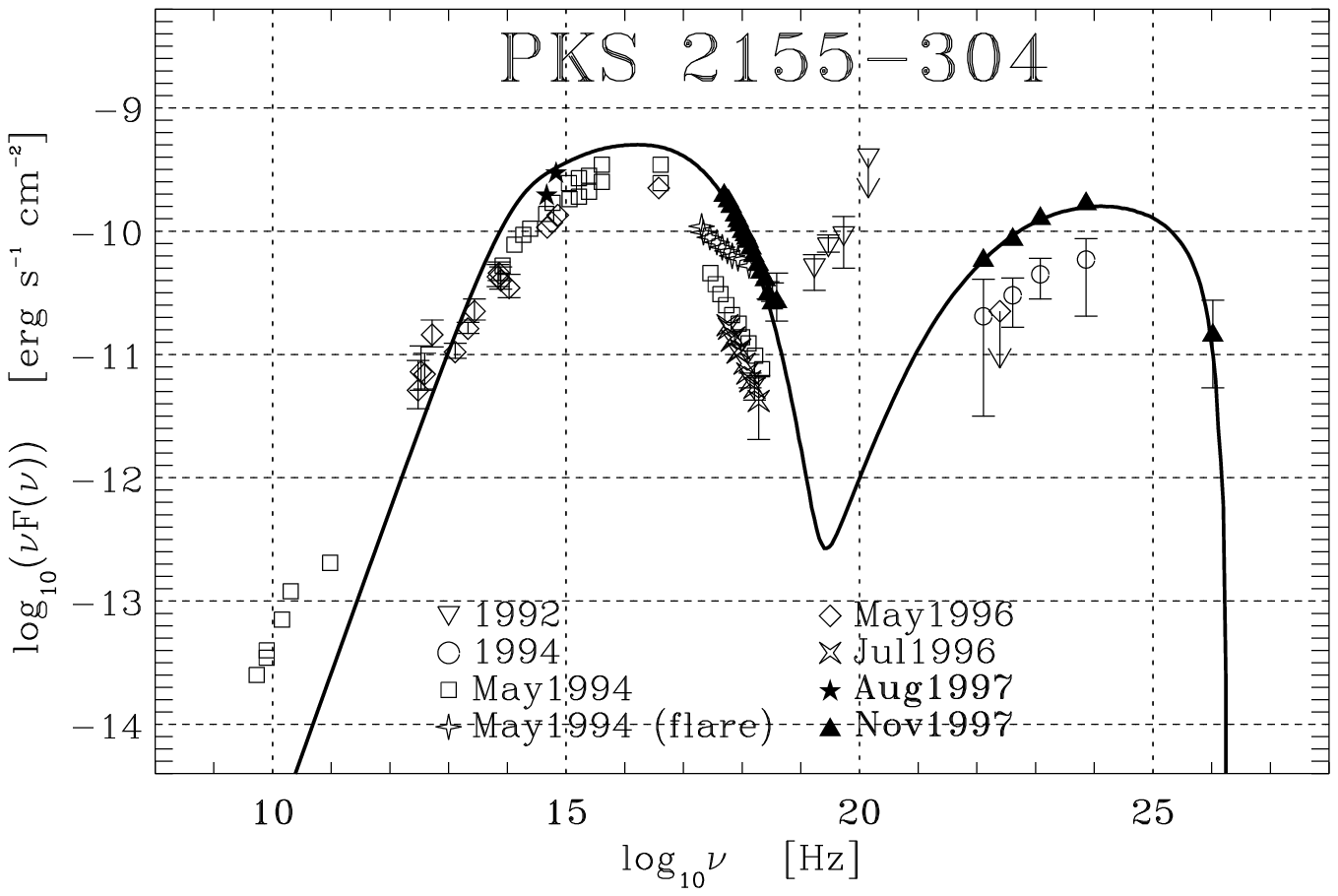}
 \epsfysize=4.5cm \hspace{-0.2cm} \epsfbox{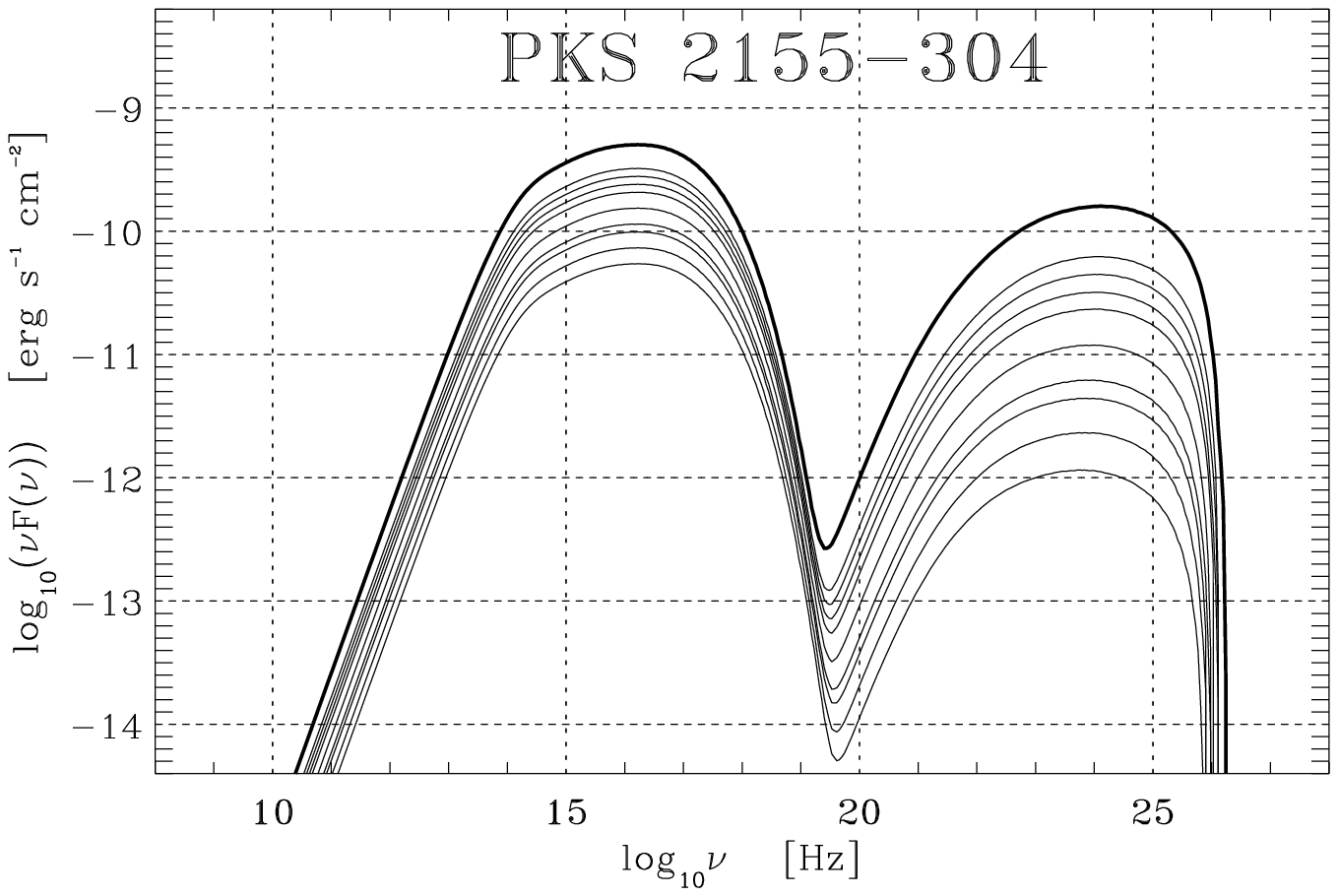}
\caption[h]{Attempts to fit the SED of PKS 2155-304, using the
datat August-November 1997. In the right panel the temporal
evolution.}\label{fig3} \vspace{-2mm}
\end{figure}
%
\begin{table}[h!]
\caption{Values of the parameters used in the simulations.
$\gamma_{min}$, $\gamma_{max}$ are the energy limits of the
injected electron distribution. $\gamma_{break}$ is the energy at
the knee and $\alpha_{1}$, $\alpha_{2}$ the power indexes of the
injected broken power law. $R$ is the size of the flaring knot
embedded in a tangled magnetic field of mean intensity $B$ (in
Gauss). $\mathcal{D}$ is the Doppler beaming factor of the knot.
$t_{esc}/t_{cr}$ is the ratio between the electrons escape time
and light-crossing time. The number density was taken the same for
the three fits: $N_{e} = 10^6$ [cm$^{-3}$]. Fig. 1 represents the
fit of May 1994 multiwavelength campaign
(radio-nearIR-optical-IUE-ASCA-EGRET), where the lower SED in the
left panel is the evolved temporal stage, of the upper SED, at
$t=2.2t_{cr}$. Fig. 2 reports the fit of May 1996 campaign
(ISO-optical-EUVE-EGRET) and July 1996 ASCA data, where the lower
SED in the left panel is the evolved temporal stage, of the upper
SED, at $t=1.6t_{cr}$. Fig. 3 represents the fit of August 1997
optical data and November 1997 campaign (RXTE-EGRET-Mark6).}
\label{tab1} \centering \vspace{2mm}
\begin{tabular}{cccccccc}
\hline \hline  \vspace{-3mm} \\
Figure& $\gamma_{min}$-$\gamma_{max}$ & $\gamma_{break}$ &
$\alpha_{1}$-$\alpha_{2}$ & B [G] &  R [cm]  & $t_{esc}/t_{cr}$ &
 $\mathcal{D}$ \\
   & $(\times 10^4)$  & $(\times 10^4)$ &  &  &  $(\times 10^{16})$ &   \vspace{1mm}  \\
 \hline \hline  \vspace{-3mm} \\
\textbf{1}~(May94) & 0.065-1000  & 2.8 & 2-2.5 & 0.2 & 1.6 & 1.8 & 30  \\
 \hline \vspace{-3mm} \\
 \textbf{2}~(May96) & 0.001-100  & 3.5 & 1.85-2.2 & 0.45 & 4.9 & 1.8 & 16 \\
 \hline \vspace{-3mm} \\
\textbf{3}~(Aug97) & 0.14-80  & 1 & 2.3-2.4 & 0.55 & 1.15 & 1.3 & 25  \\
 \hline \hline
\end{tabular}
\end{table}
\par We considered a distribution of relativistic electrons confined
by a tangled magnetic field $\mathbf{B}$, into a non-thermal
flaring knot of plasma (of dimension $R$) in the jet of PKS
2155-304. The globule is subordinated to the injection of shocked
and freshly energetic electrons, radiatively cooling. This can be
described with an appropriate truncation of the exact kinetic
relations, adopting a diffusion approximation that leads to a
random walk in the energy space of electrons. Hence we used a mere
one-dimensional diffusion-advection-like kinetic equation (Ciprini
2002) for the electron number density $N_{e}(\gamma,t)$
[cm$^{-3}$], in order to describe the time-dependent evolution of
the electron distribution into the emitting volume. The solutions
of this type of equations, are very sensitive to initial
conditions, boundary conditions, to injection and losses forms.
The overall dynamics adopted is similar to other previous works:
(e.g. Kataoka et al. 2000, Li \& Kusunose 2000, Chiaberge \&
Ghisellini 1999, Kirk et al. 1998).\par The energy distribution
injected, was assumed as a stationary broken power law
(acceleration via a two-step process or a composite electron
population), between $\gamma_{min}$ and $\gamma_{max}$, with an
high energy exponential cutoff (roll-off of the energy
distribution), in the form:
$Q_{e}(\gamma)=Q_{0}~\gamma^{-\alpha}~e^{-\gamma/\gamma_{max}}$ .
The particles population, cools by synchrotron and IC scattering.
Time light travel effects have been taken in account, for the
convolution of the spectra produced at different distances in the
source. An uniform tangled magnetic field of intensity $B$ was
considered, even if a small-amplitude plasma turbulence (e.g.
B\"{o}ttcher et al. 1997) could be superimposed. The ensemble
synchrotron spectrum was integrated by the calculated electron
distribution at any given time, and then Comptonized by the
interaction with the energetic electron distribution. Finally the
produced spectra were transformed to the observer, through the
Doppler boosting factor $D=\left[(1+z)\Gamma(1-\beta \cos
\theta)\right]^{-1}$ of the flaring blob (where $\Gamma$ is the
bulk Lorentz factor and $\theta$ the angle respect to the
observer).\par In Figures 1, 2 and 3, are represented our three
fit attempts, with their temporal evolution. The parameters, with
their values, are listed in Table 1. The observed spectra of PKS
2155-304 was simulated quite well at high energies. The radio flux
is believed to originate also from stationary, different and far
components, respect to the rapid variable flaring knot modelled,
doing the discrepancy.

\section{Conclusions}
The values of the parameters are in agreement with previous models
(e.g. Kataoka et al. 2000). The dimension of the flaring region
$R$ is on the order of $10^{16}$ cm, the magnetic field between
$0.2-0.6$ G, and a relatively high Doppler factor between $15-30$
seems characterize this TeV blazar. The monotonic course with the
frequency of the various X-ray spectra, suggest a small
variability in the synchrotron peak frequency, and indeed in our
simulations the synchrotron peak is always in the UV bands. This
model seems to be a feasible tool, to study the global energetics
and the rapid high energy variability, not only in PKS 2155-304,
but in each blazar dominated by the SSC process.

\end{document}